\documentclass[intlimits,twoside,a4paper]{article}

\usepackage{amsmath,amssymb}
\usepackage{graphicx}

\usepackage[T2A]{fontenc}
\usepackage[cp1251]{inputenc}

\usepackage[eqsecnum]{cmpj2}
\issue{2016}{19}{4}{43702}
\doinumber{10.5488/CMP.19.43702}

\title[Nature of extrinsic and intrinsic self-trapping of charge carriers]%
{Nature of extrinsic and intrinsic self-trapping of charge carriers in underdoped cuprate high-$T_\text{c}$ superconductors}%

\author[O.K. Ganiev]{O.K. Ganiev}
\address{Department of Physics, National University of Uzbekistan, 100174
Tashkent, Uzbekistan }

\authorcopyright{O.K. Ganiev, 2016}
\date{Received August 1, 2016, in final form September 17, 2016}

\begin{document}

\maketitle

\begin{abstract}
Nature of extrinsic and intrinsic self-trapping (ST) of charge
carriers in cuprates have been studied theoretically. The binding
energies and radii of the extrinsic and intrinsic large polarons
and bipolarons in cuprates are calculated variationally using the
continuum model and adiabatic approximation. We have shown that
the extrinsic and intrinsic three-dimensional (3D) large
bipolarons exist in underdoped cuprates at
$\eta=\varepsilon_{\infty}/\varepsilon_0<0.127$ and $\eta<0.138$,
respectively [where $\varepsilon_{\infty}$ ($\varepsilon_0$) is
the optic (static) dielectric constant].
\pacs 71.38.-k, 74.72.-h
\end{abstract}

\section{Introduction}

It is well known that electron-phonon coupling (EPC) is one of the most
common and fundamental interactions in solids. In particular, a
strong EPC in high-temperature superconducting cuprates (HTSC) was
reported by a lot of experiments, which implies that EPC plays an
important role in pairing. Accordingly, charge carriers in polar
materials interacting with the phonon field can cause the
formation of self-trapped polarons and bipolarons. The interest to
the polarons and bipolarons is caused by their important role in
explaining many characteristics of HTSC compounds (see \cite{Salje2005,Devreese&Alexandrov2009} and references therein).
The ground state of an undoped HTSC is an antiferromagnet (AF)
whose doping by holes leads to the phase showing high temperatures
of the superconducting (SC) transition. The extended
$t$-$t'$-$t''$-$J$-Hubbard model, where a hole in a two-dimensional (2D)
AF is simultaneously magnetic and a lattice polaron that moves
through the lattice emitting and absorbing magnons and phonons, is
used in many papers for calculation of spectral properties of
underdoped HTSC compounds \cite{Mishchenko2011}. According to the
band calculations and angle resolved-photoemission spectroscopy
(ARPES) data \cite{Imada1998,Damascelli2003}, the electronic
structure of the parent cuprate compounds is well described by a
three-band Hubbard model and the oxygen valence band lies within
the Mott-Hubbard gap. Furthermore, according to the combined study of the
ARPES and quantum Monte Carlo simulations, there was proposed a novel
polaronic metallic state in underdoped cuprates
\cite{Mishchenko2011}. The variety of opinions range from a
complete negation of the role of EPC in the physics of HTSC
\cite{Andreson1997,Andreson2007} to the statement that the binding
energy of the polaron is an order of magnitude larger than any
characteristic energy of the magnetic subsystem, and, therefore,
exactly the magnetic system is irrelevant
\cite{Alexandrov2007,Alexandrov1996}. Both of the above radical
statements have been criticized many times. On the other hand, the
success of the extended $t$-$J$-Hubbard model in describing the ARPES
\cite{Mishchenko2004,Mishchenko2011} and optical spectra
\cite{Mishchenko2006} does not make it possible to completely disregard the role of the
magnetic subsystem. Correspondingly, there is an
opinion that interactions with both magnetic and lattice
subsystems are important \cite{Mishchenko2009}. Results indicating
an important role of the EPC in HTSC compounds are presented,
with emphasis on its implications for ARPES and optical
conductivity~\cite{Mishchenko2009}.

Hole doping of the cuprates produces first quasi-free holes having
the mass $m_{\text{h}}$ in the oxygen valence band. The hole carriers are
assumed to be within both a three-dimensional (3D) and a 2D deformable medium, the
last one being CuO$_2$ layers \cite{Kastner1998}. In reality,
however, no systems can be purely 2D, and therefore, the layered
cuprate compounds may be approximated as a 3D deformable medium.
There is also a convincing experimental evidence that the
consideration of cuprates as 3D systems may appear to be more
appropriate (see \cite{Lee2006,Lavrov1998,Komiya2002,Hussey1998}.
The experimental results presented in \cite{Komiya2002} indeed
confirm that the hole-doped system La$_{2-x}$Sr$_x$CuO$_4$ (LSCO)
becomes less 2D in the strongly localized state. In polar
materials, the hole carriers interacting both with lattice
vibrations (i.e., acoustic and optical phonons) and with lattice
defects (e.g., dopants or impurities), can easily be self-trapped
near the defects and in a defect-free deformable lattice.
Actually, the carrier localization in the cuprates is interrelated
and the quantitative theory of this phenomenon is still lacking.
Particularly, possible roles of large- and small-radius dopants,
dopant-driven and carrier-driven inhomogeneities,
carrier–defect–lattice and carrier–lattice interactions and other
factors are very important for the localization of carriers in
hole-doped cuprates. In the present paper we study the extrinsic
(defect-assisted) and intrinsic (phonon-assisted) single particle
and pair self-trapping (ST) of carriers using the continuum model of ionic crystal
and adiabatic approximation. The possibility of the formation of
localized extrinsic and intrinsic (bi)polaronic states as well as
hydrogen-like impurity states in charge-transfer (CT) gap of the
cuprates is examined and compared with the experimental data.

\section{Calculation of the ground-state energy of the system of a defect-bound hole carrier in the polar crystal}

Electron-phonon interaction (EPI) affects the electronic properties
of semiconductors and polar crystals in various ways depending on
the strength of the electron-phonon interaction. Among them, the
polaron formation and the ST are its typical and important effects
on the carriers. The relevant charge carriers in hole-doped
cuprates are large polarons \cite{Kastner1998,Emin2013} and the
strong EPIs are responsible for enhancement of the polaron mass
$m_{\text{p}}=(2.0{-}3.0)m_{\text{h}}$ \cite{Basov2005} (where $m_{\text{h}}=m_{\text{e}}$ is the
free electron mass). According to Toyozawa \cite{Toyozawa1983}, the
mechanisms for ST of carriers are classified as intrinsic and
extrinsic ones. The intrinsic mechanism means that carriers are
self-trapped at deformed lattice sites through EPI. The extrinsic
mechanism consists of short-range and/or long-range potentials by
impurities or defects enhancing ST of carriers due to EPI. So far,
there are no detailed quantitative studies of the intrinsic and
extrinsic ST of the carriers in lightly doped cuprates. In order to better
understand the situation, the possibility of the formation of
extrinsic self-trapped states and of the intrinsic ones in the CT
gap of the cuprates need to be thoroughly examined and compared
with the existing experimental results confirming the existence of
such localized in-gap states.

We use a continuum model proposed in \cite{Toyozawa1983} and
adiabatic approximation to calculate the ground-state
energy of an interacting system of a defect (dopant)-bound hole
carrier in the polar crystal. In the continuous approximation, the
functionals of the total energies of the single-carrier and
two-carrier systems can be written as follows:
\begin{eqnarray}\label{1Eq}
E_1[\psi(r)]& =& \frac{\hbar^2}{2m^\ast}\int[\nabla\psi(r)]^2\rd^3r-\frac{e^2}{2\tilde\varepsilon}\int\frac{\psi^2(r)\psi^2(r')}{|r-r'|}\rd^3r\,\rd^3r'\nonumber\\
&& - \frac{E_{\text{d}}^2}{2K}\int\psi^4(r)\rd^3r-\frac{Ze^2}{\varepsilon_0}\int\frac{\psi^2(r)}{r}\rd^3r +
\left(V_0-\frac{E_{\text{d}}E_{\text{dD}}}{K}\right)\int\psi^2(r)\delta(r)\rd^3r
\end{eqnarray}
and
\begin{eqnarray}\label{2Eq}
E_2[\Psi(r_1,r_2)]&=&\frac{\hbar^2}{2m^\ast}\int\left[|\nabla_1\Psi(r_1,r_2)|^2+|\nabla_2\Psi(r_1,r_2)|^2\right]\rd^3r_1\,\rd^3r_2
+\frac{e^2}{\varepsilon_{\infty}}\int\frac{\Psi^2(r_1,r_2)}{|r_1-r_2|}\rd^3r_1\,\rd^3r_2\nonumber\\
&&-\frac{2e^2}{\tilde\varepsilon}\int\frac{\Psi^2(r_1,r_2)\Psi^2(r_3,r_4)}{|r_1-r_3|}\rd^3r_1\,\rd^3r_2\,\rd^3r_3\,\rd^3r_4\nonumber\\
&&-\frac{2E_{\text{d}}^2}{K}\int\Psi^2(r_1,r_2)\Psi^2(r_2,r_3)\rd^3r_1\,\rd^3r_2\,\rd^3r_3
-\frac{2Ze^2}{\varepsilon_0}\int\frac{\Psi^2(r_1,r_2)}{r_1}\rd^3r_1\,\rd^3r_2\nonumber\\
&&+2\left(V_0-\frac{E_{\text{d}}E_{\text{dD}}}{K}\right)\int\Psi^2(r_1,r_2)\delta(r_1)\rd^3r_1\,\rd^3r_2\,,
\end{eqnarray}
where $m^*$  and $e$ are carrier effective mass and charge,
respectively, $r_1$ and $r_2$ are the position vectors of the
carriers, $E_{\text{d}}$ and $E_{\text{dD}}$ are the deformation potentials of the
carrier and the defect, respectively, $\tilde\varepsilon$ is given
by
$\tilde\varepsilon^{-1}=\varepsilon^{-1}_{\infty}-\varepsilon^{-1}_0=(1-\eta)/\varepsilon_{\infty}\,$,
$K$ is an elastic constant, $V_0$ is the short-range defect
potential, $Z$ is the charge state of the defect. A large ionicity
of the cuprates $\eta=\varepsilon_{\infty}/\varepsilon_0\ll1$,
enhances the polar EPI and the tendency to polaron formation.

In order to minimize the functionals (\ref{1Eq}) and (\ref{2Eq})
with respect to $\psi(r)$ and $\Psi(r_{1},r_{2})$ we can take the
trial wave functions as follows:
\begin{equation}\label{3Eq}
\psi(r)=N_1\exp\left[-(\sigma r)\right]
\end{equation}
and
\begin{equation}\label{4Eq}
\Psi(r_1,r_2)=N_2[1+\gamma(\sigma r_{12})]\exp\left[-\sigma
(r_1+r_2)\right],
\end{equation}
where $N_1=\sigma^{3/2}/\sqrt{\pi}$ and
$N_2=\sigma^3/\pi\sqrt{K_1(\gamma)}$ are the normalization
factors, $\sigma=\beta/a_0$,
$K_1(\gamma)=1+\frac{35}{8}\gamma+6\gamma^2$ is the correlation
coefficient, $\beta$ and $\gamma$ are the variational parameters that
characterize the localization degree of carriers and the
correlation in their motions, respectively. $r_{12}=|r_1-r_2|$ is
the distance between the carriers, $a_0$ is the lattice constant.
Substituting equations (\ref{3Eq}) and (\ref{4Eq}) into equations
(\ref{1Eq}) and (\ref{2Eq}), and performing the integrations in
equations (\ref{1Eq}) and (\ref{2Eq}), we obtain the following
functionals
\begin{eqnarray}\label{5Eq}
E_1[\beta]=A\left[\beta^2-g_{\text{s}}(1+b_{\text{s}})\beta^3-g_{\,\text{l}}\left(1-\eta+\frac{16}{5}Z\eta\right)\beta\right]
\end{eqnarray}
and
\begin{eqnarray}\label{6Eq}
E_2[\beta,\gamma]&=&2A\frac{K_2(\gamma)}{K_1(\gamma)}\left\{\beta^2-
\frac{8}{5}g_{\,\text{l}}\right.\left[2(1-\eta)\frac{K_4(\gamma)}{K_1(\gamma)K_2(\gamma)}
+Z\eta\frac{K_5(\gamma)}{K_2(\gamma)}-\frac{K_3(\gamma)}{K_2(\gamma)}\right]\beta\nonumber\\
&&\left.-16g_{\text{s}}\left[\frac{K_6(\gamma)}{K_1(\gamma)K_2(\gamma)}+\frac{b_{\text{s}}}{16}\frac{K_7(\gamma)}{K_2(\gamma)}\right]\beta^3\right\},
\end{eqnarray}
where $A$=$\hbar^2/2m^*a^2_0$, $g_{\text{s}}$=$E^2_{\text{d}}/16\pi Ka_0^3A$ and
$g_{\,\text{l}}=5e^2/16\varepsilon_{\infty}a_0A$ are dimensionless
short-range and long-range carrier-phonon coupling parameters,
$b_{s}=16\left(E_{\text{dD}}/E_{\text{d}}-KV_0/E^2_{\text{d}}\right)$ is short-range
carrier-defect-pho\-non coupling parameter, and the correlation
coefficients are analytical functions of the variational parameter~$\gamma$:
\begin{align}
&K_2(\gamma)=1+\frac{25}{8}\gamma+4\gamma^2,\nonumber\\
&K_4(\gamma)=\frac{5}{8}+\frac{1087}{216}\gamma+\frac{38237}{2304}
\gamma^2+\frac{67639}{2592}\gamma^3+\frac{4293}{256}\gamma^4,\nonumber\\
&K_3(\gamma)=\frac{5}{8}\gamma+2\gamma+\frac{35}{16}\gamma^2,\qquad K_5(\gamma)=2+\frac{15}{2}\gamma+9\gamma^2,\nonumber\\
&K_6(\gamma)=\frac{1}{8}+\frac{185}{216}\gamma+\frac{4199}{1728}
\gamma^2+\frac{8591}{2592}\gamma^3+\frac{477}{256}\gamma^4,\nonumber\\
&K_7(\gamma)=1+3\gamma+3\gamma^2.\nonumber
\end{align}
Minimization of the functionals (\ref{5Eq}) and (\ref{6Eq}) over
the variational parameters $\beta$ and $\gamma$ would give the
ground state energies $E_1^{\text{min}}(\beta)$ and
$E_2^{\text{min}}(\beta,\gamma)$ of single and pair self-trapped
carriers. As a function of $\beta$ and $\gamma$, these functionals
have also got the maxima $E_1^{\text{max}}(\beta)$ and
$E_2^{\text{max}}(\beta,\gamma)$. The other parameters $A$, $g_{\text{s}}$, $g_{\,\text{l}}$,
$b_{\text{s}}$ and $Z$ entering equations~(\ref{5Eq}) and (\ref{6Eq}) play
different roles in the formation of self-trapped states and
determine the nature of the localized states of carriers in doped
polar materials. The values of the parameters $A$, $g_{\text{s}}$ and $g_{\,\text{l}}$
can be obtained using the experimental values of the parameters
$a_0$, $m^*$, $K$, $\varepsilon_{\infty}\,$, $\varepsilon_0$ and the
Fermi energy $E_\text{F}$ of the undoped cuprates. The parameters $b_{\text{s}}$
and $Z$ characterize the formation of extrinsic (i.e.,
hydrogenic and non-hydrogenic) self-trapped states of carriers,
whereas the parameters $g_{\text{s}}$ and $g_{\,\text{l}}$ characterizing the
strengths of short- and long-range carrier-phonon interactions
are responsible for the formation of intrinsic self-trapped
states.

Using the equations (\ref{5Eq}) and (\ref{6Eq}), we calculate the
energies of different localized in-gap states in the CT gap of
the cuprates. In order to determine the nature of these in-gap
states and the quasi-free to localized state transition, we
distinguish different physical situations in these systems. One
can make interesting analyses of real systems based on the
sign and magnitude of $b_{\text{s}}$. In hole-doped cuprates, the
situations might be quite different for different types of
dopants. In particular, the signs of the deformation potential
constants $E_{\text{d}}$ and $E_{\text{dD}}$ for holes and small-radius defects
are always positive, while $E_{\text{dD}}$ for large-radius defects is
negative \cite{Toyozawa1983}. However, at
present no information is available for the magnitudes of the parameters
$E_{\text{dD}}$ and $V_0$. Therefore, the parameter $b_{\text{s}}$ in equations~(\ref{5Eq}) and
(\ref{6Eq}) can be considered as the free parameter. We consider
first the possibility of the formation of localized in-gap states at
single and pair ST of carriers near the small-radius dopants (with
$E_{\text{dD}}>0$ or $b_{\text{s}}>0$) in La-based cuprates. In this case, both
short and long range parts of the defect potential in
equations (\ref{1Eq}) and (\ref{2Eq}) are attractive, so that the
substitution of small-radius cations (e.g., $\rm{Ca^{2+}}$ and
$\rm{Nd^{3+}}$ ions) for $\rm{La^{3+}}$ ions in
$\rm{La_{2}CuO_{4}}$ and for $\rm{Sr^{2+}}$ ions in LSCO leads to
a combined defect- and phonon-assisted ST of hole carriers with
the formation of localized single-carrier and two-carrier
impurity states, which are extrinsic polaronic and bipolaronic
(the so-called U-pairing) states. At $Z\neq0$ and $b_{\text{s}}>0$, the minima
of $E_1(\beta)$ and $E_2(\beta,\gamma)$ correspond to the
ground-state energies of the extrinsic large polaron and
bipolaron, respectively, measured with respect to the top of the
oxygen valence band. The binding energies of such extrinsic large
polaron and bipolaron (or negative U center) are defined as
$E_{\text{pI}}=|E_{1}^{\text{min}}(\beta)|$ and
$E_{\text{bU}}=|E_{2}^{\text{min}}(\beta,\gamma)-2E_1^{\text{min}}(\beta)|$,
respectively. In 3D systems, there is generally a potential barrier
between the large- and small-radius self-trapped states. The two
states of the extrinsic large polaron are separated by a potential
barrier, with activation energy
$E^{\text{A}}_1=E_1^{\text{max}}(\beta)-E_1^{\text{min}}(\beta)$ needed for the
transition from the large-radius localized state to the
small-radius one. The potential barrier
$E^{\text{A}}_2=E_2^{\text{max}}(\beta,\gamma)-E_2^{\text{min}}(\beta)$ exists between
the large and small-radius extrinsic bipolaronic states.

\begin{figure}[!t]
 \centering
\includegraphics[width=0.48\textwidth]{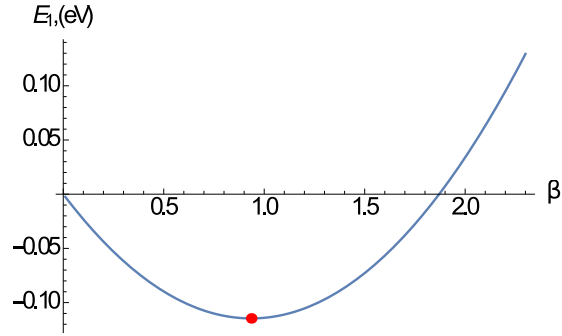}
\caption{\label{fig1} (Color online) The dependence of
ground-state energy of extrinsic large polaron on the variational
parameter $\beta$ for $\eta=0.08$ in La-based cuprates. The red
point indicates a single carrier self-trapped state.}
\end{figure}

\begin{figure}[!t]
 \centering
\includegraphics[width=0.63\textwidth]{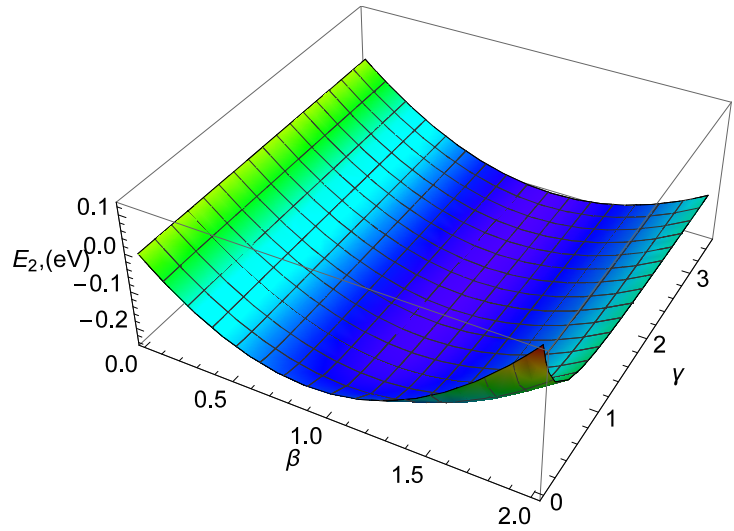}
\caption{\label{fig2} (Color online) The dependence of
ground-state energy of extrinsic large bipolaron on the
variational parameters $\beta$ and $\gamma$ for $\eta=0.08$ in
La-based cuprates.}
\end{figure}

We now calculate the basic parameters of the extrinsic large
(bi)polarons in La-based cuprates. At low temperature, the
$\rm{La}$-based cuprates are orthorhombic with the lattice
parameter $a_0\simeq5.4$~{\AA}. According to the spectroscopy data,
the Fermi energy of the undoped cuprates is about $E_{\text{F}}\simeq7$~eV
\cite{Lu1990}. To determine the value of the short-range
carrier-phonon coupling constant $g_{\text{s}}$, we can estimate the
deformation potential $E_{\text{d}}$ as $E_{\text{d}}=(2/3)E_{\text{F}}$ \cite{Kittel1967}.
For the cuprates, typical values of other parameters are $m^*=m_{\text{e}}$
\cite{Kastner1998}, $\varepsilon_{\infty}=3{-}5$
\cite{Emin1989,Weger1994}, $K=1.4\cdot10^{12}$~dyn/cm$^2$
\cite{Baetzold1990}, and $Z=1$. The calculated values of $E_{\text{pI}}$,
$E_{\text{bU}}$, $E_1^{\text{A}}$ and $E_2^{\text{A}}$ for $b_{\text{s}}=1$ and different values of
$\varepsilon_{\infty}$ and $\eta$ are presented in table
\ref{table1}. From table \ref{table1} we can see that the potential barriers
separating the large- and small-radius extrinsic (bi)polaronic
states are rather high. These high potential barriers prevent the
formation of small extrinsic (bi)polarons in 3D cuprates. The
defect- and phonon-assisted ST of large polaron and large
bipolaron in La-based cuprates are shown in figures \ref{fig1} and
\ref{fig2}, respectively.
\begin{table*}[!h]
\caption{\label{blobs} The calculated parameters of
the extrinsic large polarons and bipolarons (with correlation
between the pairing  carriers) in 3D cuprates at $Z=1$, $b_{\text{s}}=1$
and different values of $\varepsilon_{\infty}$ and $\eta$.}
\vspace{2ex}
\centering \scriptsize
\begin{tabular}
{|p{11pt}|p{21pt}|p{23pt}|p{21pt}|p{21pt}|p{21pt}|p{23pt}|p{21pt}|p{21pt}|p{21pt}|p{23pt}|p{21pt}|p{21pt}|}
\hline\hline \raisebox{-1.50ex}[0cm][0cm]
{\centerline{$\eta$}} &
\multicolumn{4}{|p{100pt}|}{\centerline{$\varepsilon_{\infty}=3.5$}}
&
\multicolumn{4}{|p{100pt}|}{\centerline{$\varepsilon_{\infty}=4$}}
&
\multicolumn{4}{|p{100pt}|}{\centerline{$\varepsilon_{\infty}=4.5$}} \\
\cline{2-13}
 &
$E_{\text{pI}}$,\,eV & $E_{\text{bU}}$,\,eV & $E_1^{\text{A}}$,\,eV & $E_2^{\text{A}}$,\,eV& $E_{\text{pI}}$,\,eV &
$E_{\text{bU}}$,\,eV & $E_1^{\text{A}}$,\,eV & $E_2^{\text{A}}$,\,eV & $E_{\text{pI}}$,\,eV & $E_{\text{bU}}$,\,eV
& $E_1^{\text{A}}$,\,eV &
$E_2^{\text{A}}$,\,eV \\
 \hline \hline
 0.00 & 0.1135 &0.0610 &5.3373 &5.8611
 &0.0863 &0.0456 &5.6962 &6.4752
 &0.0679 &0.0354 &5.9807 &6.7958
 \\
\hline 0.02 &0.1240 &0.0525 &5.2128 &5.7649
 &0.0943 &0.0391 &5.5849 &6.3881
 &0.0741 &0.0303 &5.8801 &6.7090
 \\
\hline 0.04 &0.1349 &0.0434 &5.0893 &5.6693
 &0.1026 &0.0321 &5.4742 &6.3013
 &0.0806 &0.0247 &5.7800 &6.6224
 \\
\hline 0.06 &0.1464 &0.0336 &4.9668 &5.5743
 &0.1113 &0.0246 &5.3643 &6.2150
 &0.0874 &0.0188 &5.6804 &6.5359
 \\
\hline 0.08 &0.1584 &0.0231 &4.8453 &5.4799
 &0.1203 &0.0166 &5.2552 &6.1292
 &0.0945 &0.0125 &5.5815 &6.4495
 \\
\hline 0.10 &0.1709 &0.0120 &4.7248 &5.3861
 &0.1298 &0.0081 &5.1468 &6.0438
 &0.1019 &0.0058 &5.4831 &6.3633
 \\
\hline 0.12 &0.1839 &0.0001 &4.6053 &5.2929
 &0.1396 &\centering{---} &5.0391 &5.9588
 &0.1096 &\centering{---} &5.3854 &6.2773
  \\
\hline\hline
\end{tabular}
\label{table1}
\end{table*}

Another interesting question is in what way large-radius dopants in
cuprates affect the carrier-pho\-non system, especially near such
defects. This opposite situation is realized in
LSCO or {La$_{2-x}$Ba$_x$CuO$_4$}
(LBCO), where the radius of $\rm{Sr^{2+}}$ ions is larger than
that of $\rm{La^{3+}}$ ions \cite{Markert1990}, so that for
Sr$^{2+}$ ion $Z=1$, $E_{\text{dD}}<0$ or $b_{s}<0$. In this case, the
short-range part of the impurity potential in equation~(\ref{1Eq}) is
repulsive. Therefore, one can treat it like a hard core. The
hole-lattice interactions near the large-radius dopants in LSCO
and LBCO are suppressed by this repulsive defect potential and
hole carriers are localized at a distance from the dopants (i.e.,
hole-carriers are loosely bound to dopants by long-range Coulomb
attraction). From these considerations, it follows that the
hole-lattice interaction near the large-radius dopants is weak and
the localized impurity state may be of a hydrogen-like character
described by a rigid lattice model \cite{Mott1974}. Therefore, we
can consider the hydrogen-like impurity centers having the Bohr
radius $a_{\text{H}}=0.529\varepsilon_{0}(m_{\text e}/m^{*})$~{\AA} and the
ionization energy $E^{\text H}_{\text{I}}=e^{2}/2\varepsilon_{0}a_{\text{H}}$ in lightly
doped LSCO and LBCO. A similar argument can be made for other
cuprates, such as $\rm{La_2CuO_{4+\delta}}$,
$\rm{YBa_2Cu_3O_{7-\delta}}$ (YBCO) and
$\rm{Bi_2Sr_2CaCu_2O_{8+\delta}}$ (Bi-2212). In these systems, the
doping centers are excess oxygen atoms which capture the electrons
from the oxygen conduction band and form the acceptor centers,
i.e., $\rm{O^-}$ ions. One can assume that such extra $\rm{O^-}$
ions just like $\rm{Sr^{2+}}$ ions may have negative $E_{\text{dD}}$.

\begin{figure}[!t]
 \centering
\includegraphics[width=0.67\textwidth]{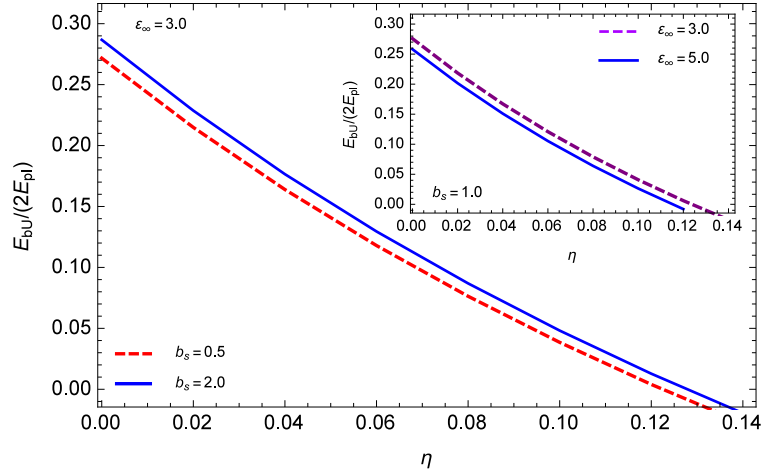}
\caption{\label{fig3} (Color online) The ratio of the binding energy
of the extrinsic large bipolaron to twice the extrinsic large
polaron binding energy as a function of $\eta$ for two values of
$b_{\text{s}}$ in 3D cuprates. The inset shows the dependence of the ratio
of the binding energy of the extrinsic large bipolaron to twice
that of the extrinsic large polaron on $\eta$ for two values of
$\varepsilon_\infty$ in 3D cuprates.}
\end{figure}

Figure \ref{fig3} shows the variation of the ratio
$E_{\text{bU}}/2E_{\text{pI}}$ with $\eta$ for $b_{\text{s}}=0.5$ and 2.0 for the
stability region of the extrinsic large bipolaron in 3D cuprates.
One can find out that such bipolarons exist as long as $\eta$ is less
than the critical value $\eta_{\text{c}}=0.127$ and the ratio
$E_{\text{bU}}/2E_{\text{pI}}$ reaches up to 0.287 (at $\varepsilon_\infty=3$
and $\eta\rightarrow0$). We have determined the stability region
of the extrinsic large bipolaron in cuprates and found that such
bipolarons exist as long as $\eta$ is less than the critical value
$\eta_{\text{c}}=0.127$ and the ratio $E_{\text{bU}}/2E_{\text{pI}}$ reaches up to 0.287
(at $\varepsilon_{\infty}=3$ and $\eta\rightarrow0$).
\begin{figure}[!b]
 \centering
\includegraphics[width=0.67\textwidth]{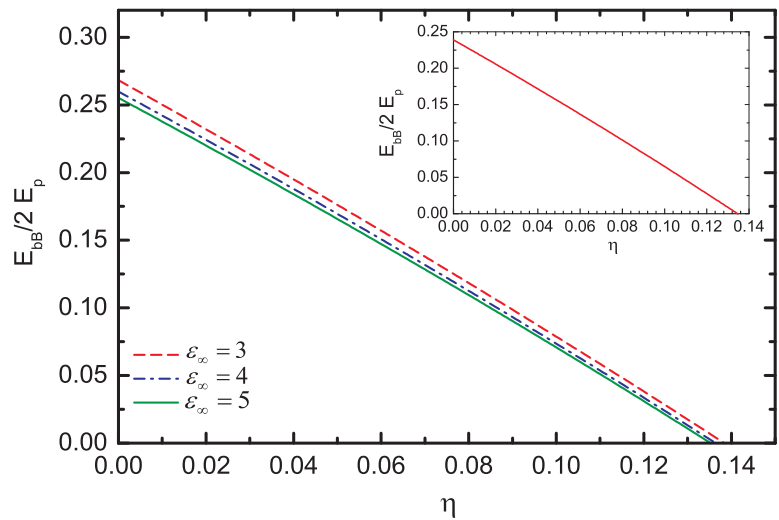}
\caption{\label{fig4} (Color online) The ratio of the binding energy
of the real large bipolaron to twice that of the real large
polaron as a function of $\eta$ for different values of
$\varepsilon_\infty$ in 3D cuprates. The inset illustrates the
ratio of the binding energy of the optical large bipolaron to
twice that of the optical large polaron on $\eta$ for
$\varepsilon_\infty=3.0$ in 3D cuprates.}
\end{figure}
Figure \ref{fig4} shows the variation of the ratio $E_{\text{bB}}/2E_{\text{p}}$
with $\eta$ for $\varepsilon_{\infty}=3$, 4 and 5 for real large
bipolarons in cuprates. One can see that in 3D cuprates, the
intrinsic large bipolarons can exist at $\eta=\eta_{\text{c}}\leqslant0.138$ and
the ratio $E_{\text{bB}}/2E_{\text{p}}$ reaches up to 0.27 (at
$\varepsilon_{\infty}=3$ and $\eta\rightarrow0$). While the inset
in figure \ref{fig4} shows the variation of the ratio
$E_{\text{bB}}/2E_{\text{p}}$ [calculated from equations (\ref{5Eq}) and
(\ref{6Eq}) at $g_{\text{s}}=0$ and $Z=0$] with $\eta$ for large optical
bipolaron. Of course, besides the ratio
$\eta=\varepsilon_{\infty}/\varepsilon_0$, the values of the
Fr\"{o}hlich electron-phonon coupling constant $\alpha$ are very
important for the formation of large optical bipolarons. The
long-range coupling of carriers with optical phonons is much
stronger than their short-range coupling with acoustic phonons.
Therefore, the long-range Fr\"{o}hlich-type EPI in polar materials
have been studied extensively \cite{Appel1975,Verbist1991},
although the short-range deformation potential type interaction is
also important and leads to new effects. The dimentionless EPI
Fr\"{o}hlich coupling constant is defined as
\begin{eqnarray}\label{07Eq}
\alpha=\frac{e^2(1-\eta)}{2\varepsilon_{\infty}\hbar\omega_{\text{LO}}}\left(\frac{2m^*\omega_{\text{LO}}}{\hbar}\right)^{1/2},
\end{eqnarray}
where $\omega_{\text{LO}}$ is the frequency of the longitudinal-optical
(LO) phonon in an ionic crystal.

In polar materials, the formation of optical bipolarons is favored
by larger values of $\alpha$ and by smaller values of $\eta$
\cite{Devreese&Alexandrov2009,Verbist1991}, i.e., the optical
(Fr\"{o}hlich or Pekar) bipolarons exist only if $\alpha$ is is
greater than a critical value $\alpha_{\text{c}}$ and when $\eta<\eta_{\text{c}}$.
Such 3D bipolarons can exist above rather high critical values
$\alpha_{\text{c}}$, e.g., $\alpha_{\text{c}}=7.3$ as found by Adamowski
\cite{Adamowski1989} and $\alpha_{\text{c}}=6.8$ found by Verbist, Peeters
and Devreese \cite{Verbist-PRB}. Further, the value of $\alpha_{\text{c}}$
corresponding to the onset of the strong coupling regime is found
to be $\alpha_{\text{c}}=5.8$ \cite{Baimatov1997}, below which the
formation of large optical bipolaron in 3D systems is unlikely. At
a given value of $\eta$, the value of $\alpha_{\text{c}}$ depends on $m^*$,
$\varepsilon_{\infty}$ and $\omega_{\text{LO}}$. The values of
$\hbar\omega_{\text{LO}}$ in high-$T_{\text{c}}$ cuprates range from 0.03 to 0.05~eV \cite{Kastner1998,Thomsen1990}. Then, according to equation
(\ref{07Eq}), the values of $\alpha$ corresponding to these
high-$T_{\text{c}}$ materials with $m^*=m_{\text{e}}$, $\varepsilon_{\infty}=3$ and
$\eta=0.02{-}0.10$ are equal to $\alpha=4.96{-}6.95$. Thus, the
conditions for the formation of large optical bipolarons are more
favorable in the cuprates with $\varepsilon_{\infty}=3$,
$\eta=0.02{-}0.06$ and $\hbar\omega_{\text{LO}}=0.03{-}0.04$~eV, at which
conditions $\alpha>\alpha_{\text{c}}\simeq5.8$ \cite{Baimatov1997} and
$\alpha>\alpha_{\text{c}}=6.8$ \cite{Verbist-PRB} are well satisfied.

We note here that the largest values of $\eta_{\text{c}}=0.079{-}0.14$ and
$E_{\text{bB}}/2E_{\text{p}}=0.22{-}0.25$ as found in the literature
\cite{Suprun1982,Hiramoto1985,Vinetskii1989} (see also
\cite{Verbist1991}) were obtained for the optical bipolaron and do
not correspond to specific substances. At the same time, the critical values
of $\eta$ below which the optical bipolarons can exist in cuprates
were small enough ($\eta_{\text{c}}=0.040{-}0.055$), as estimated in
\cite{Verbist1991}. Our results are quite impressive in the sense
that both the real and the optical bipolaron in 3D cuprates, can
really exist for relatively large values of $\eta$ (figure
\ref{fig4}) and the large bipolarons are formed with the binding
energies reaching up to 27\% (at $\eta\rightarrow0$) of twice the
large polaron energy. The distinctive feature of the cuprates is
their very large ratio of static to high-frequency dielectric
constants. This situation is favorable for carriers attracted to
polarization well created by the other ones or to Coulomb centers
(dopants) to form 3D intrinsic or extrinsic large bipolarons. At
$\varepsilon_0>30$, such bipolarons (pair states) can be formed in
lightly doped cuprates and they become unstable in an underdoped
regime. As can be seen from tables \ref{table1} and \ref{table2},
the binding energies of extrinsic and intrinsic large (bi)polarons
and the ratios $E_{\text{bU}}/2E_{\text{pI}}$ and $E_{\text{bB}}/2E_{\text{p}}$ depend on
several parameters. In particular, $E_{\text{pI}}$, $E_{\text{p}}$, $E_{\text{bU}}$ and
$E_{\text{bB}}$ would rapidly increase with
$\varepsilon_{\infty}$ decreasing from 5 to 3, while the ratios
$E_{\text{bU}}/2E_{\text{pI}}$ and $E_{\text{bB}}/2E_{\text{p}}$ increase more slowly as
$\varepsilon_{\infty}$ decreases. Interestingly, $E_{\text{pI}}$ is an
increasing function of $\eta$ (table \ref{table1}), while $E_{\text{p}}$ is
a decreasing function of $\eta$ (table \ref{table2}). Furthermore,
we find that both $E_{\text{bU}}$ and $E_{\text{bB}}$ decrease with
increasing $\eta$. We also find that the ratio $E_{\text{bU}}/2E_{\text{pI}}$
appreciably increases with $b_{\text{s}}$ as shown in figure \ref{fig3}.
The calculated values of the radii of extrinsic large (bi)polarons
$R_{\text{I}}$ $(R_{\text{BI}})$ (at $Z=1$ and $b_{\text{s}}=1$) and intrinsic large
(bi)polarons, $R_{\text{p}}$ ($R_{\text{B}}$) (at $Z=0$ and $b_{\text{s}}=0$) for different
values of $\varepsilon_\infty$ and $\eta$ are presented in table~\ref{table3}.
\begin{table*}[!h]\caption{Calculated parameters of
the real large polarons and bipolarons (with correlation between
the pairing  carriers) in 3D cuprates at different values of
$\varepsilon_{\infty}$ and $\eta$.}
\vspace{2ex}\centering \scriptsize
\begin{tabular}
{|p{11pt}|p{21pt}|p{23pt}|p{21pt}|p{21pt}|p{21pt}|p{23pt}|p{21pt}|p{21pt}|p{21pt}|p{23pt}|p{21pt}|p{21pt}|}
\hline\hline \raisebox{-1.50ex}[0cm][0cm]{\centerline{$\eta$}} &
\multicolumn{4}{|p{100pt}|}{\centerline{$\varepsilon_{\infty}=3.5$}}
&
\multicolumn{4}{|p{100pt}|}{\centerline{$\varepsilon_{\infty}=4$}}
&
\multicolumn{4}{|p{100pt}|}{\centerline{$\varepsilon_{\infty}=4.5$}} \\
\cline{2-13} & $E_{\text{p}}$,\,eV &$E_{\text{bB}}$,\,eV & $E_{\text{p}}^{\text{A}}$,\,eV & $E_{\text{B}}^{\text{A}}$,\,eV&
$E_{\text{p}}$,\,eV & $E_{\text{bB}}$,\,eV & $E_{\text{p}}^{\text{A}}$,\,eV & $E_{\text{B}}^{\text{A}}$,\,eV & $E_{\text{p}}$,\,eV &
$E_{\text{bB}}$,\,eV &
$E_{\text{p}}^{\text{A}}$,\,eV &$E_{\text{B}}^{\text{A}}$,\,eV \\
\hline\hline 0.00 &0.1107 &0.0583 &27.275 &19.989
 &0.0845 &0.0439 &28.049 &21.063
 &0.0666 &0.0343 &28.656 &21.744
 \\
\hline 0.02 &0.1063 &0.0484 &27.398 &20.269
 &0.0811 &0.0364 &28.158 &21.313
 &0.0640 &0.0284 &28.753 &21.980
 \\
\hline 0.04 &0.1019 &0.0389 &27.521 &20.551
 &0.0778 &0.0293 &28.267 &21.564
 &0.0614 &0.0228 &28.851 &22.217
 \\
\hline 0.06 &0.0977 &0.0299 &27.645 &20.834
 &0.0746 &0.0225 &28.376 &21.815
 &0.0588 &0.0175 &28.949 &22.453
 \\
\hline 0.08 &0.0935 &0.0215 &27.769 &21.118
 &0.0714 &0.0161 &28.485 &22.068
 &0.0563 &0.0125 &29.046 &22.690
 \\
\hline 0.10 &0.0895 &0.0136 &27.893 &21.404
 &0.0683 &0.0101 &28.595 &22.321
 &0.0539 &0.0078 &29.144 &22.928
 \\
\hline 0.12 &0.0855 &0.0061 &28.017 &21.691
 &0.0653 &0.0044 &28.704 &22.576
 &0.0515 &0.0003 &29.242 &23.166
 \\
\hline 0.14 &0.0816 &\centering{---} &28.142 &21.980
 &0.0623 &\centering{---} &28.814 &22.832
 &0.0491 &\centering{---} &29.341 &23.404
  \\
\hline\hline
\end{tabular}
\label{table2}
\end{table*}
\begin{table*}[!h]
\caption{Calculated values of the radii of intrinsic and extrinsic
large (bi)polarons $R_{\text{p}}$, $R_{\text{I}}$, $R_{\text{B}}$ and $R_{\text{BI}}$ in 3D cuprates
at different values of $\varepsilon_\infty$ and $\eta$.}
\vspace{2ex}
\centering \scriptsize
\begin{tabular}
{|p{11pt}|p{21pt}|p{23pt}|p{21pt}|p{21pt}|p{21pt}|p{23pt}|p{21pt}|p{21pt}|p{21pt}|p{23pt}|p{21pt}|p{21pt}|}
\hline\hline \raisebox{-1.50ex}[0cm][0cm]{\centerline{$\eta$}} &
\multicolumn{4}{|p{100pt}|}{\centerline{$\varepsilon_{\infty}=3.5$}}
&
\multicolumn{4}{|p{100pt}|}{\centerline{$\varepsilon_{\infty}=4$}}
&
\multicolumn{4}{|p{100pt}|}{\centerline{$\varepsilon_{\infty}=4.5$}} \\
\cline{2-13}
 &
$R_{\text{p}}$, {\AA} & $R_{\text{B}}$, {\AA} & $R_{\text{I}}$, {\AA} & $R_{\text{BI}}$,
{\AA} &
$R_{\text{p}}$, {\AA} & $R_{\text{B}}$, {\AA} & $R_{\text{I}}$, {\AA} & $R_{\text{BI}}$,
{\AA} &
$R_{\text{p}}$, {\AA} & $R_{\text{B}}$, {\AA} & $R_{\text{I}}$, {\AA} & $R_{\text{BI}}$, {\AA} \\
\hline\hline 0.00 &8.6096 &13.045 &8.2839 &12.403
 &9.8840 &15.057 &9.5633 &14.434
 &11.158 &17.066 &10.841 &16.457
 \\
\hline 0.02 &8.7917 &13.577 &7.9061 &12.084
 &10.092 &15.663 &9.1322 &14.072
 &11.392 &17.746 &10.357 &16.052
 \\
\hline 0.04 &8.9813 &14.146 &7.5586 &11.777
 &10.309 &16.311 &8.7357 &13.724
 &11.636 &18.475 &9.9111 &15.662
 \\
\hline 0.06 &9.1791 &14.756 &7.2378 &11.482
 &10.535 &17.007 &8.3699 &13.390
 &11.890 &19.256 &9.5001 &15.288
 \\
\hline 0.08 &9.3854 &15.413 &6.9409 &11.199
 &10.770 &17.756 &8.0313 &13.068
 &12.155 &20.098 &9.1197 &14.928
 \\
\hline 0.10 &9.6009 &16.121 &6.6652 &10.926
 &11.017 &18.564 &7.7169 &12.758
 &12.432 &21.005 &8.7665 &14.581
 \\
\hline 0.12 &9.8261 &16.887 &6.4084 &10.662
 &11.274 &19.438 &7.4243 &12.460
 &12.722 &21.986 &8.4379 &14.248
 \\
\hline 0.14 &10.062 &17.718 &6.1688 &10.408
 &11.543 &20.386 &7.1512 &12.173
 &13.025 &23.051 &8.1312 &13.926
 \\
\hline\hline
\end{tabular}
\label{table3}
\end{table*}

\section{Discussion}

We now make some remarks about the characteristic (i.e., binding)
energies of large bipolarons in the cuprates. The binding energies
of extrinsic and intrinsic bipolarons strongly depend on
$\varepsilon_{\infty}\,$, $\varepsilon_0$ and $\eta$. The values of
$\varepsilon_{\infty}$ and $\varepsilon_0$ observed in high-$T_{\text{c}}$
cuprates are 3--5 \cite{Imada1998,Tranquada1995,Zaanen1998} and
33--105 \cite{Imada1998,Varyukhin1981}, respectively, so that the
values of $\eta$ range from 0.028 to 0.15. Using the values of
$\varepsilon_{\infty}=3$ and $\eta\simeq0.03$, we find
$E_{\text{bU}}\simeq0.07$~eV and $E_{\text{bB}}\simeq0.061$~eV. If we take other
experimental values of $\varepsilon_{\infty}=4$ and
$\varepsilon_0\simeq50$ for the cuprates (see \cite{Devreese&Alexandrov2009,Verbist1991}), then we obtain
$E_{\text{bU}}\simeq0.017$~eV and $E_{\text{bB}}\simeq0.016$~eV at $\eta=0.08$.
Further, at $\varepsilon_{\infty}=5$ and $\eta=0.08$ we find
$E_{\text{bU}}\simeq0.0097$~eV and $E_{\text{bB}}\simeq0.0102$~eV. Thus, the
extrinsic and intrinsic bipolarons can be experimentally found in
high-$T_{\text{c}}$ cuprates in the energy ranges $\sim0.01{-}0.07$~eV and
$\sim0.01{-}0.06$~eV, respectively. The binding energies of large
polarons and bipolarons are manifested in the excitation spectra
of the hole-doped cuprates as the temperature-independent
low-energy gaps or pseudogaps, which are different from the
high-energy CT gaps ($\Delta_{\text{CT}}\simeq1.5{-}2.0$~eV
\cite{Kastner1998}) of the cuprates.

It is of interest to compare our results with experimental data on
localized in-gap states (or bands) and energy gaps (which are
responsible for the existance of insulating phase and are
precursors to the pseudogaps observed in the metallic state) in
hole-doped cuprates. The above extrinsic and intrinsic
(bi)polaronic states as well as hydrogenic impurity states emerge
in the CT gap of the cuprates. In the experiments, these localized
states are displayed as the in-gap states. One can see that the
value of $E_{\text{pI}}\simeq0.13$~eV obtained at
$\varepsilon_{\infty}=4$ and $\eta=0.1$ (table~\ref{table1}) is
consistent with experimental data for lightly doped
$\rm{La_2CuO_{4+\delta}}$ \cite{Kastner1998}. The in-gap impurity
band observed in this system at 0.13~eV might be associated with
the extrinsic large polarons. While the values of
$E_{\text{p}}\simeq0.096{-}0.105$~eV (table~\ref{table2}) obtained at
$\varepsilon_{\infty}=3.5$ and $\eta$=0.04--0.06 agree reasonably
well with the large pseudogap value $\sim$0.1~eV observed in LSCO
\cite{Ino1998}. One of the important experimental observations is
that in LSCO, the flatband \cite{Ino2002}, which is $\sim$0.12~eV
below the Fermi energy for $x=0.05$, moves upwards monotonously
with increasing $x$, but the flatband is lowered as $x$ decreases
and loses its intensity in the insulating phase. Apparently, the
flatband observed by ARPES in the lightly doped LSCO ($x=0.05$) is
the energy band of large polarons since the effective mass of
carriers obtained from the analysis of the ARPES spectra is about
2.1$m_{\text{e}}$ \cite{Ino2002}. The values of $R_{\text{p}}$ (table~\ref{table3})
are also in good agreement with the experimental values of the
radii of polarons which vary from 6 to 10~{\AA} in cuprates
\cite{Kastner1998}.

\section{Conclusions}

We have studied the possible mechanisms of carrier localization in
inhomogeneous hole-doped cup\-rates. The quantitative theory of the
impurity-assisted and phonon-assisted single particle and pair ST
of hole carriers in 3D lightly doped cuprates is developed within
the continuum model and adiabatic approximation. The possible
mechanisms for carrier localization lead to the formation of
extrinsic large (bi)polaronic states, the hydrogenic impurity
states (i.e., impurities with loosely bound free carriers or large
polarons) and intrinsic large (bi)polaronic states in the CT
gap of the cuprates. We have variationally calculated the binding
energies and radii of the extrinsic and intrinsic large polarons
and bipolarons, taking into account the short- and long-range
parts of the carrier-defect-phonon and carrier-phonon
interactions. We have determined the stability region of the
extrinsic large bipolaron in cuprates and found that such
bipolarons exist as long as $\eta$ is less than the critical value
$\eta_{\text{c}}=0.127$ and the ratio $E_{\text{bU}}/2E_{\text{pI}}$ (where $E_{\text{pI}}$ and
$E_{\text{bU}}$ are the binding energies of the extrinsic large polaron
and bipolaron, respectively) reaches 0.287 (at
$\varepsilon_\infty=3$ and $\eta\rightarrow0$). We have obtained
the conditions for the real large bipolaron stability and
estimated the values of $E_{\text{bB}}/2E_{\text{p}}$ in 3D cuprates (where
$E_{\text{p}}$ and $E_{\text{bB}}$ are the binding energies of the intrinsic
large polaron and bipolaron, respectively).

\section*{Acknowledgements}
This work was supported by the Foundation of National University
of Uzbekistan under Grant No: F2-FA-F115.

\newpage
\ukrainianpart

\title{Природа зовнішніх і внутрішніх самозахопних носіїв струму в слабкозлегованих купратних високотемпературних напівпровідниках}%

\author{O.K. Ганєв}
\address{Фізичний факультет, Університет Узбекистану, 100174 Ташкент, Узбекистан }

\makeukrtitle

\begin{abstract}
Теоретично вивчено природу зовнішніх і внутрішніх самозахопних носіїв струму в  купратах.
Енергії зв'язування і радіуси зовнішніх і внутрішніх великих поляронів і біполяронів в купратах обчислено варіаційно, використовуючи
неперервну модель та адіабатичне наближення. Ми показали, що зовнішні і внутрішні тривимірні великі біполярони існують в слабозлегованих
купратах, відповідно,  при
$\eta=\varepsilon_{\infty}/\varepsilon_0<0.127$ і $\eta<0.138$ [де $\varepsilon_{\infty}$ ($\varepsilon_0$) --- оптична (статична)
діелектрична стала].
\keywords полярон, біполярон, самозахоплення, високотемпературні надпровідники
\end{abstract}
\end{document}